\documentclass[%
 aip,
 amsmath,amssymb,
 reprint,%
]{revtex4-1}
\usepackage{graphicx}
\usepackage{dcolumn}
\usepackage[table]{xcolor}
\usepackage{bm}
\usepackage[utf8]{inputenc}
\usepackage[T1]{fontenc}
\usepackage{mathptmx}

\begin{document}

\preprint{AIP/123-QED}

\title{Geometrical magnetoresistance effect and mobility in graphene field-effect transistors}

\author{Isabel Harrysson Rodrigues}
\email[]{isabelr@chalmers.se}
\affiliation{Department of Microtechnology and Nanoscience, Chalmers University of Technology, SE-41296 Gothenburg, Sweden}
\author{Andrey Generalov}
\affiliation{VTT Technical Research Center of Finland Ltd., P.O. Box 1000, FI-02044 VTT, Finland}
\author{Anamul Md Hoque}
\affiliation{Department of Microtechnology and Nanoscience, Chalmers University of Technology, SE-41296 Gothenburg, Sweden}
\author{Miika Soikkeli}
\affiliation{VTT Technical Research Center of Finland Ltd., P.O. Box 1000, FI-02044 VTT, Finland}
\author{Anton Murros}
\affiliation{VTT Technical Research Center of Finland Ltd., P.O. Box 1000, FI-02044 VTT, Finland}
\author{Sanna Arpiainen}
\affiliation{VTT Technical Research Center of Finland Ltd., P.O. Box 1000, FI-02044 VTT, Finland}
\author{Andrei Vorobiev}
\affiliation{Department of Microtechnology and Nanoscience, Chalmers University of Technology, SE-41296 Gothenburg, Sweden}

\date{\today}

\begin{abstract}
Further development of the graphene field-effect transistors (GFETs) for high-frequency electronics requires accurate evaluation and study of the mobility of charge carriers in a specific device. Here, we demonstrate that the mobility in the GFETs can be directly characterized and studied using the geometrical magnetoresistance (gMR) effect. The method is free from the limitations of other approaches since it does not require an assumption of the constant mobility and the knowledge of the gate capacitance. Studies of a few sets of GFETs in the wide range of transverse magnetic fields indicate that the gMR effect dominates up to approximately 0.55~T. In higher fields, the physical magnetoresistance effect starts to contribute. The advantages of the gMR approach allowed us to interpret the measured dependencies of mobility on the gate voltage, i.e., carrier concentration, and identify the corresponding scattering mechanisms. In particular, the range of the fairly constant mobility is associated with the dominating Coulomb scattering. The decrease in mobility at higher carrier concentrations is associated with the contribution of the phonon scattering. Analysis shows that the gMR mobility is typically 2-3 times higher than that found via the commonly used drain resistance model. The latter underestimates the mobility since it does not take the interfacial capacitance into account.
\end{abstract}


\maketitle
    Future progress in modern electronics relies on the development of novel two-dimensional materials with cutting-edge performance, among which graphene is a promising candidate. The very high carrier mobility and velocity in graphene could enable much faster electronics than traditional semiconductors. The room-temperature intrinsic mobility in single layer graphene is above 100 000~cm$^2$/Vs, which is larger than that in the highest mobility III-V compounds.\cite{Chen2008, DasSarma2011, Mayorov2011} With such high mobility, the graphene-based high-frequency electronics might reach the still uncovered terahertz range offering many exciting novel applications.\cite{Schwierz2013IEEE} However, currently, in real graphene devices, mobility is significantly reduced. In particular, the room temperature mobility in the graphene field-effect transistors (GFETs), with the highest reported high-frequency performance, is below 5000~cm$^2$/Vs.\cite{Bonmann2019,Asad2021} The mobility degradation is associated with material imperfections caused by the specific device processing and vicinity of dielectrics in the device structure.\cite{Bonmann2017} Additionally, there is typically a strong surface distribution of the mobility measured in the GFETs located at different positions on the wafer caused by the spatially inhomogeneous Coulomb potential associated with charged impurities.\cite{Adam18392,Asad2020} Therefore, for further development of the graphene-based high-frequency electronics, methods of accurate evaluation of the mobility directly based on the measured characteristics of the specific device, i.e., without involving different test structures, should be developed and applied.
\par
    Up to date, the mobility in a specific GFET is generally characterized using a drain resistance model applied to the measured transfer characteristics.\cite{Meric2008, kim2009letter, Zhong2015} This approach does not require additional test structures. However, it has a number of limitations including the assumption of constant mobility and uncertainty of the gate capacitance, which can be strongly modified by the interfacial states.\cite{bonmann2017olof} This may result in large errors in the mobility evaluation. For example, we have shown, that the mobility values calculated using the drain resistance model can be 2-3 times lower than those found from the delay time analysis in the same GFETs.\cite{Bonmann2017} Already in the early years of the development of the field-effect transistors, it has been proposed that the carrier mobility and velocity can be assessed through the geometrical magnetoresistance (gMR) effect.\cite{Masselink1985,Masselink1986} This effect arises when the magnetic field causes the path of the charge carriers to deviate from a straight line, raising the sample resistance.\cite{BOOKschroeder} It was indicated that advantages of the gMR method of evaluation of the mobility, in comparison with other methods, are that it does not require knowledge of the carrier concentration or the transistor's capacitance, gate length, access resistance, and threshold voltage.\cite{Sakowicz2006} In our previous work, we have studied the low-field mobility and high-field carrier velocity in InGaAs/InP high-electron-mobility transistors found via the gMR effect.\cite{HarryssonRodrigues2022} In this work, we demonstrate that mobility in GFETs can be directly characterized using the gMR method. This method is free from the limitations of the drain resistance approach since it does not require an assumption of a constant mobility, or knowledge of the gate capacitance. 
\par
    The paper is structured in two parts. The first part focuses on demonstrating the gMR effect in the GFETs and evaluating the gMR mobility and the associated charge carrier transport mechanisms. The second part reports on a comparative analysis of the mobility applying the commonly used drain resistance model approach, which shows that it may underestimate the mobility 2-3 times because of uncertainty in the gate capacitance.
\par
\begin{figure}[ht]
            \centering
            \includegraphics[width=\columnwidth]{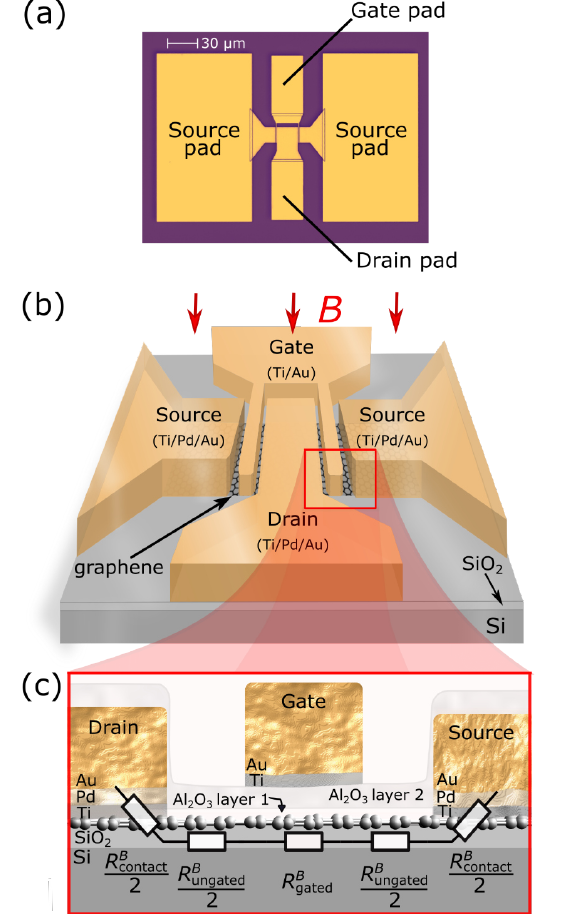}
            \caption{(a) A typical optical microscopy image and (b) a 3D illustration of the GFET with transverse $B$-field applied, together with (c) a schematic cross-sectional view of the active area indicating the materials of different layers. Shown also on (c) is the total resistance equivalent circuit at the condition of applied $B$-field. Note that the first and second gate dielectric layers, labeled as Al$_2$O$_3$ layer 1 and Al$_2$O$_3$ layer 2, respectively, are assumed to be transparent and, hence not visible in (b).}
            \label{fig:device}\label{fig:1abc}
        \end{figure}

    The GFETs studied in this work have layouts similar to those previously published,\cite{Bonmann2019} i.e., with dual gate-fingers centered between the source and drain contacts, with 100~nm long ungated regions, as illustrated in Fig.~\ref{fig:1abc}. Two sets of GFETs with the gate length ($L$) and width ($W$) varying in the ranges of 0.2-2.0~$\mu$m and 5-15~$\mu$m, respectively, were fabricated on two different Si wafers, at VTT and Chalmers, which will be referred to below as wafer-1 and wafer-2, respectively. The processing steps are similar to those described previously.\cite{Asad2020}
\par    
    The four main distinguishable stages (i-iv) of the GFET fabrication are as follows. In stage (i), the chemical vapor deposition (CVD) graphene film, prepared by Graphenea, is transferred onto high resistivity silicon/silicon oxide (Si/SiO$_2$) substrate, with a SiO$_2$ thickness of 1~$\mu$m, using the Easy Transfer approach.~\cite{graphenea} The transferred graphene film is covered by an approximately 8~nm thick Al$_2$O$_3$ layer constituting the first layer of the gate dielectric. This layer is obtained by six times repeating the steps of the deposition of 1~nm thick Al film and its subsequent oxidation on a hotplate at 160$^o$~C for 5~minutes. In this technology, the first gate dielectric layer encapsulates graphene, preventing contamination during further processing. In stage (ii), the graphene/dielectric mesa is patterned with e-beam lithography, followed by an etch of the Al$_2$O$_3$ layer using the buffered oxide etch diluted by 10 parts of water (BOE/H$_2$O) and a subsequent etch of the graphene using oxygen plasma. The openings in the Al$_2$O$_3$ layer for the drain/source contacts are patterned with e-beam lithography followed by an etch using BOE/H$_2$O. The drain/source contacts are formed by deposition of 1~nm Ti/15~nm Pd/250~nm Au layered structure and the use of standard lift-off process. In stage (iii), the second gate dielectric layer of Al$_2$O$_3$ is formed by repeating the process described above ten times. The second gate dielectric layer is approximately 14~nm thick making the total gate dielectric thickness of 22~nm. The second dielectric layer covers the graphene edges exposed at the mesa sidewalls and, hence, prevents short-circuiting by the overlapping gate fingers. In stage (iv), the gate electrodes and the contact pads are fabricated by e-beam lithography and deposition of 10~nm Ti/290~nm Au layered structure by e-beam evaporation followed by a standard lift-off process. Fig.~\ref{fig:1abc}~(a) shows a typical optical microscope image of a fabricated GFET. 
\par    
    As shown and discussed below, despite the similarity of the processing steps, the series resistance in the GFETs on wafer-1 is typically more than 10 times larger than that on wafer-2. However, the gMR mobility can be extracted after de-embedding the series resistance. This allows us to demonstrate that the gMR method is applicable on GFETs with significantly different series resistance and mobility values.
\par
    The transfer characteristics of the GFETs were measured in the one-finger, common source configuration at the drain voltage $V_{DS} = 0.3$~V and $-0.1$~V on wafer-1 and wafer-2, respectively, using a Keithley 2612B dual-channel source meter, as well as a HP 4156B semiconductor parameter analyzer without and with a transverse magnetic field. The measurements were performed in a standard Hall set-up equipped with a movable permanent magnet with magnetic flux density ($B$-field) $B=0.33$~T, see figure in the supplementary material. Several GFETs have been measured with the aim of confirming the gMR effect and making a comprehensive error analysis, see the supplementary material. To further verify the gMR effect, the dependencies of the drain resistances on the transverse $B$-field in the range of 0-0.8~T, in both directions, were measured under vacuum at room temperature using a current source (Keithley 6221), a nanovoltmeter (Keithley 2182A) and an electromagnet.
\par    
    The measurements of the transfer characteristics without a magnetic field were done comparatively in the dark and illuminated environment to investigate variations of the drain resistance, which were shown to remain within the errors of measurements. This allowed for excluding the possible effects of the persistent photo-conductivity traps,\cite{Masselink1986} caused by screening the light by the magnet, and confirms that the charge carrier transport under magnetic field is governed mainly by the gMR effect.
\par
    Fig.~\ref{fig:1abc}~(b) and (c) show a 3D illustration of the GFET with transverse $B$-field applied, together with a schematic cross-sectional view of the active area indicating materials of different layers indicated. Fig.~\ref{fig:1abc}~c) also shows the equivalent circuit of the total drain-source resistance under an applied magnetic field $B$. The $R^B_{\mathrm{gated}}$ and $\frac{1}{2}R^B_{\mathrm{ungated}}$ are the resistances of the gated and ungated regions of the channel, respectively. The $\frac{1}{2}R^B_{\mathrm{contact}}$ is the contact resistance associated with the graphene-metal junction. At the condition without $B$-field, the corresponding resistances are notated as $R^0_{\mathrm{gated}}$, $\frac{1}{2}R^0_{\mathrm{ungated}}$, and $\frac{1}{2}R^0_{\mathrm{contact}}$. The resistances of the ungated regions and contact resistances constitute the total series resistances as $R^B_{\mathrm{series}} = R^B_{\mathrm{ungated}} + R^B_{\mathrm{contact}}$ and $R^0_{\mathrm{series}} = R^0_{\mathrm{ungated}} + R^0_{\mathrm{contact}}$ with and without $B$-field, respectively. The total GFET resistance can be expressed as $R^B_{\mathrm{total}} = R^B_{\mathrm{series}} + R^B_{\mathrm{gated}}$ and $R^0_{\mathrm{total}} = R^0_{\mathrm{series}} + R^0_{\mathrm{gated}}$ with and without $B$-field, respectively.

\par
\begin{figure}[ht]
            \centering
            \includegraphics[width=\columnwidth]{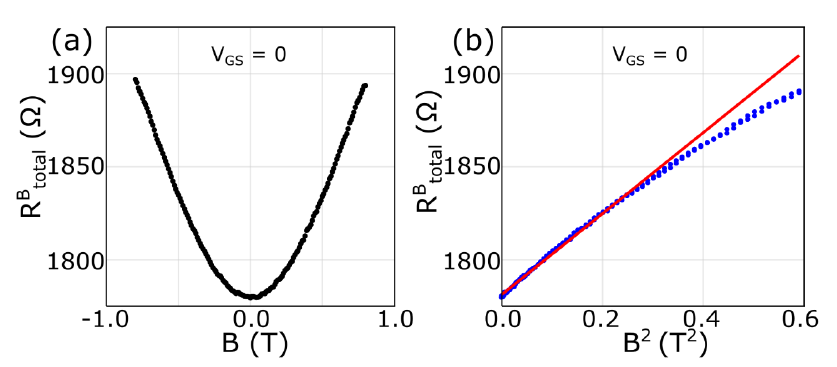}
            \caption{(a) The $R^B_{\mathrm{total}}$ of a GFET from wafer-1 versus transverse $B$-field varying in both directions at zero gate voltage. (b) The same data plotted versus the square of the $B$-field. The straight line indicates a pure quadratic dependence of $R^B_{\mathrm{total}}$ versus $B$.}
            \label{fig:2ab}
        \end{figure}

    Fig.~\ref{fig:2ab}~(a) shows a typical dependence of the $R^B_{\mathrm{total}}$, measured between the drain and source terminals, of a GFET from wafer-1, with $L=1\ \mu$m and $W=5\ \mu$m on the transverse $B$-field varied in both directions at zero gate voltage. 
    It can be seen, that the dependence is symmetric and with the drain resistance increasing with the $B$-field as a power law function. Under the conditions of a negligible physical magnetoresistance effect (pMR) and $L/W<0.4$, the gMR mobility is given by Ref.~$\cite{BOOKschroeder}$
        \begin{equation}\label{eq:u_gMR}
            \mu_{\mathrm{gMR}} \approx \frac{1}{B}\sqrt{\frac{R^B_{\mathrm{gated}}}{R^0_{\mathrm{gated}}}-1},
        \end{equation}
\par
    In general, the series resistance depends on the magnetic field due to the gMR effect in the ungated regions of the channel.\cite{HarryssonRodrigues2022} Since the top and bottom graphene-dielectric interfaces of gated and ungated regions of the channel are similar, one can assume that at zero gate voltage ($V_{GS} = 0$) the gated and ungated regions are indistinguishable. One can also assume that the variation of contact resistance due to gMR effect in the transfer area of the graphene-metal contact is negligible because of the dominating transverse electric field component and, hence, negligible Lorentz force.~\cite{Nagashio2011} Therefore, the contact resistance does not depend on the magnetic field and $R^B_{\mathrm{contact}}=R^0_{\mathrm{contact}}$. With these assumptions one can readily get from Eq.~(\ref{eq:u_gMR}) that
        \begin{equation}\label{eq:2}
            R^B_{\mathrm{total}} = (R^0_{\mathrm{total}}-R{^0_\mathrm{contact}})\big(\mu_{\mathrm{gMR}}B\big)^2 + R^0_{\mathrm{total}}.
        \end{equation}
    It can be seen from Eq.~(\ref{eq:2}), that the dependence of the $R^B_{\mathrm{total}}$ on $B^2$ should reveal a straight line. Fig.~\ref{fig:2ab}~(b) shows the same experimental data presented in Fig.~\ref{fig:2ab}~(a) plotted versus $B^2$. The dependence is fairly linear up to $B\approx0.55$~T, indicating that the approximation by Eq.~(\ref{eq:u_gMR}) is valid below this $B$-field. The sub-linear behavior at higher $B$-field is in agreement with that calculated using exact solution.~\cite{Masselink1985} This behaviour is in a qualitative agreement with a similar dependence observed on AlGaAs MODFETs.~\cite{Masselink1985} 
\par
\begin{figure}[ht]
            \centering
            \includegraphics[width=\columnwidth]{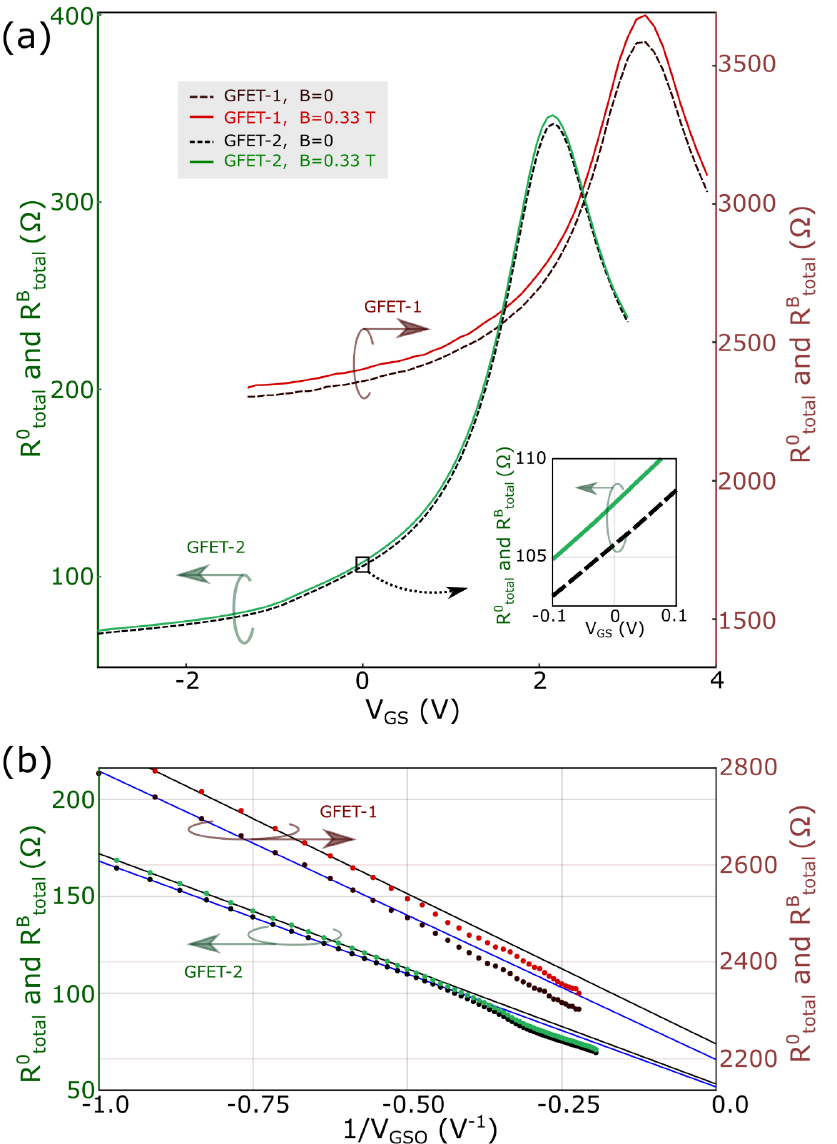}
            \caption{(a) The drain resistance of GFET-1 and GFET-2 versus gate voltage with and without the transverse $B$-field. Inset shows the narrow range of the gate voltage from $-0.1$~V to 0.1~V, for a clearer demonstration of the gMR effect, in GFET-2. (b) The drain resistances of GFET-1 and GFET-2, with and without the transverse $B$-field, plotted versus $\frac{1}{V_{GSO}}$ in the hole conductivity branch above $-1$~V$^{-1}$. The straight lines are linear fits in the ranges of constant mobility.}
            \label{fig:3}
        \end{figure}

    Fig.~\ref{fig:3}~(a) shows typical dependencies of the drain resistances on the gate voltage of a GFET on wafer-1 (referred below as GFET-1) and a GFET on wafer-2 (referred below as GFET-2), with and without transverse magnetic field. GFET-1 and GFET-2 have the same gate length of 1~$\mu$m, but different gate width of $5\ \mu$m, and $15\ \mu$m, respectively. It can be seen, that the magnetic field increases the drain resistance in the whole range of the gate voltage, apparently, due to the gMR effect, see also inset in Fig.~\ref{fig:3}~(a).
\par    
    The gMR mobility can be calculated using Eq.~(\ref{eq:u_gMR}) after subtracting the corresponding series resistances from the measured drain resistances. To find the series resistances we applied an approach similar to that published previously.\cite{Lusakowski2006} In the GFETs, at relatively high gate voltage overdrive $V_{\mathrm{GSO}} = V_{\mathrm{GS}} - V_{\mathrm{Dir}}$, where $V_{\mathrm{Dir}}$ is the Dirac voltage, i.e., under the condition of the gate induced concentration much larger than the residual concentration of the charge carriers ($n_\mathrm{0}$), the drain resistances with (and without) magnetic field can be expressed as
        \begin{equation}\label{eq:drainresistance}
            R^0_{\mathrm{total}} = R^0_{\mathrm{series}} + \frac{L}{Wen\mu},
        \end{equation}
    where $e$ is the elementary charge, $\mu$ is the field-effect mobility and $n$ is the charge carrier concentration proportional to $V_{\mathrm{GSO}}$.\cite{dorgan2010} Based on our previous studies, we assume that in the limited range of $n$ the carrier transport is governed by the Coulomb scattering with the mobility independent of the concentration.\cite{Asad2020} For comparison, analysis of the carrier density in the graphene test structures showed that mobility is relatively constant in the range of concentration 2-3$\times10^{12}$~cm$^{-2}$,~\cite{dorgan2010} which is explained by the dominating Coulomb scattering. It was shown that the Coulomb scattering is the only mechanism resulting in mobility being independent on the carrier concentration.\cite{Adam18392,dorgan2010,Giannazzo2011} Under these conditions, the drain resistances given by Eq.~(\ref{eq:drainresistance}) are linear functions of the $\frac{1}{V_{\mathrm{GSO}}}$, and the series resistances can be found by linear fitting of the corresponding dependencies. In the analysis below, we consider only hole branches of the transfer characteristics, i.e., at $V_{\mathrm{GSO}} < 0$, because the number of data in the electron branches is not sufficient for reliable fitting and the series resistance in the electron branch is typically higher due to the formation of the $pn$-junction in the ungated regions.\cite{Bidmeshkipour2015} Fig.~\ref{fig:3}~(b) shows the series resistances of GFET-1 and GFET-2 plotted versus $\frac{1}{V_{\mathrm{GSO}}}$ in the hole conductivity branch above $-1$~V$^{-1}$ together with the linear fits made in the ranges of the constant mobility. It can be seen that the dependencies are fairly linear up to approximately $-0.5$~V$^{-1}$, manifesting that the mobility is constant. Deviations from the linear dependencies above $-0.5$~V$^{-1}$, i.e., higher carrier concentration, can be explained by a decrease in mobility due to the increasing contribution of phonon scattering.\cite{dorgan2010} The linear fits give the series resistances of 2198~$\Omega$ and 52~$\Omega$, with corresponding specific-width resistivity of 5495~$\Omega\times\mu$m and 780~$\Omega\times\mu$m for GFET-1 and GFET-2, respectively. The GFETs on wafer-1 typically reveal relatively larger specific-width resistivity than the GFETs on wafer-2, which can be explained by incomplete removal of the Al$_2$O$_3$ layer in the drain/source contact openings.
\par
\begin{figure}[ht]
            \centering
            \includegraphics[width=\columnwidth]{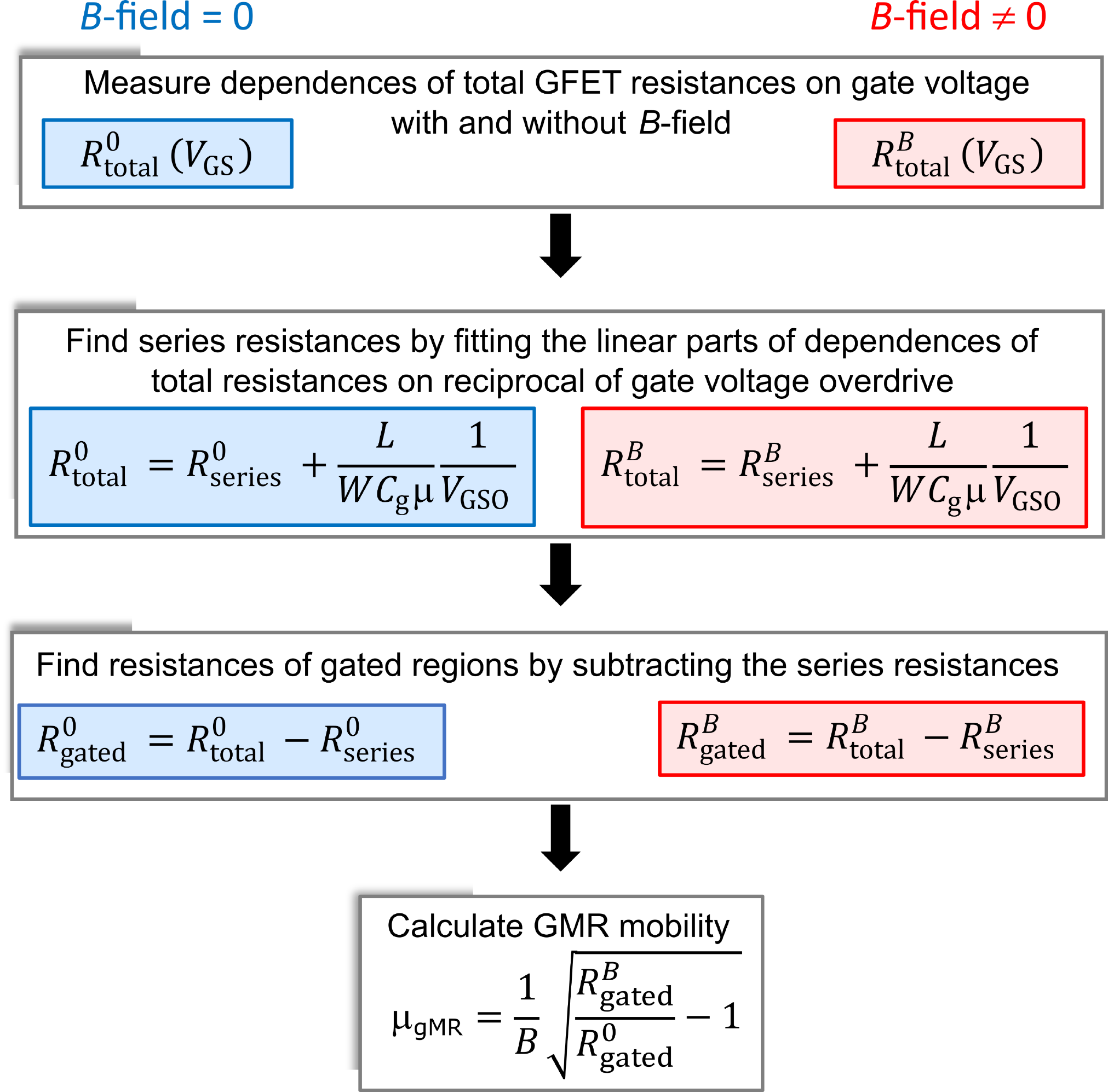}
            \caption{Flowchart showing algorithm of extraction of the gMR mobility used in this work.}
            \label{fig:flow}
        \end{figure}

    
    Fig.~\ref{fig:flow} shows a flowchart of the algorithm used in this work to extract the gMR mobility. The algorithm consists of two similar sequences of de-embedding the series resistances of a GFET without $B$-field (left part of the flowchart) and with $B$-field (right part of the flowchart) followed by calculation of the gMR mobility using Eq.~\ref{eq:u_gMR}.
\par
\begin{figure}[ht]
            \centering
            \includegraphics[width=\columnwidth]{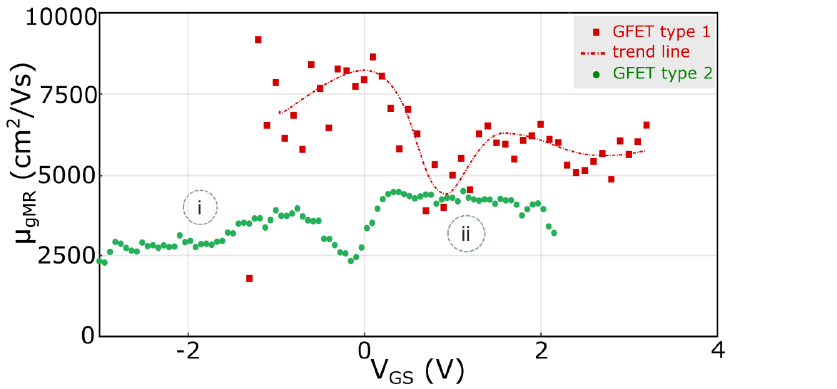}
            \caption{The gMR mobility of GFET-1 (squares) and GFET-2 (circles) versus gate voltage. Two different regions with different dominating charge carrier mechanisms are indicated as (i) and (ii) for GFET-2. The trend line is shown as a guide for the eye.}
            \label{fig:5}
        \end{figure}

    Fig.~\ref{fig:5} shows the gMR mobility calculated using Eq.~(\ref{eq:u_gMR}) with the series resistances found by the linear fits of the dependencies in Fig.~\ref{fig:3}~(b). It can be seen, that, in general, the mobility depends on the gate voltage, i.e., carrier concentration. In the dependence of GFET-2, one can distinguish two different regions. In the region (i) for $V_{\mathrm{GS}}$ up to approximately -0.5~V, the mobility decreases with increasing the gate voltage modulus, which can be associated with increasing contribution of phonon scattering at higher carrier concentrations. The decrease of mobility with carrier concentration has been previously observed and similarly explained by changing the dominant scattering mechanism from Coulomb to phonon scattering.\cite{dorgan2010} In the region (ii) of $V_{\mathrm{GS}}$ between approximately 0.5 and 2~V, the mobility is relatively constant indicating that Coulomb scattering dominates. It can be seen from Fig.~\ref{fig:5} that between the regions (i) and (ii) there is a drop in the mobility at $V_{\mathrm{GS}}\approx-0.1$~V. One can notice that this gate-source voltage coincides with the drain-source voltage bias used in the measurements of GFET-2. Therefore, the gate-drain voltage is approximately zero, and the channel potential at the gated region's drain side equals that of the ungated region. Under this condition, there is no $pp$-junction formed between the gated and ungated region, as described e.g., in Ref.~$\cite{Kim2013}$, and, hence, the series resistance is lower compared to that at other gate-source voltages below and above this value. Since, in the applied algorithm, the series resistance is found in the region (ii), this local minimum of the series resistance results in an underestimation of the actual $\mu_{\mathrm{gMR}}$, which manifests itself as a drop in mobility at $V_{\mathrm{GS}}\approx V_{\mathrm{DS}}$. As shown in Fig.~\ref{fig:5}, the data from GFET-1 is relatively more scattered than that of GFET-2, which most likely is a result of drain current instabilities caused by tunneling of the charge carriers through the Al$_2$O$_3$ residuals in the graphene-metal junctions. This complicates the physical interpretation of the dependence found for GFET-1. However, the clearly seen drop in mobility at 0.5-1.0~V approximately coincides with the drain voltage bias used in the measurements of GFET-1, which confirms our explanation above. Note, that this differs from $V_{\mathrm{DS}}=-0.1$~V used in the experiments and the drop in mobility seen in Fig.~\ref{fig:5} for GFET-2.
\par
\begin{figure}[ht]
            \centering
            \includegraphics[width=\columnwidth]{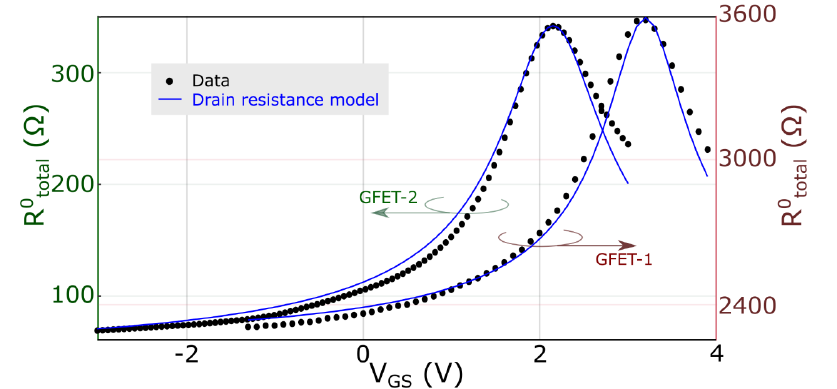}
            \caption{The drain resistances of GFET-1 and GFET-2 versus gate voltage without magnetic field. The solid lines represent the drain resistance model fitting with the commonly applied approach, i.e., using the $R^\mathrm{0}_{\mathrm{series}}$, $n_\mathrm{0}$ and $\mu$ as fitting parameters. Note that the plot aspect ratio differs from that in Fig.~\ref{fig:3}~(a) showing in part the same experimental data.}
            \label{fig:6}
        \end{figure}

    For comparison, we evaluated the field-effect mobility by applying the commonly used drain resistance model
        \begin{equation}\label{eq:4MericKim}
                R^\mathrm{0}_{\mathrm{total}} = R^\mathrm{0}_{\mathrm{series}} + \frac{L}{W}\frac{1}{\mu e}\frac{1}{\sqrt{n^2_0+\big(V_{\mathrm{GSO}}\frac{C_\mathrm{g}}{e}\big)^2}},
        \end{equation}
    where $C_\mathrm{g}$ is the gate capacitance per unit area.\cite{Meric2008, kim2009letter} It can be shown, that under the conditions used in our experiments, the graphene quantum capacitance can be ignored.\cite{kim2009letter} Fig.~\ref{fig:6} shows the $R^\mathrm{0}_{\mathrm{total}}$ versus gate voltage of GFET-1 and GFET-2, measured without $B$-field, together with fitting by the drain resistance model. The solid lines represent the best fitting using the commonly applied approach, i.e., using the $R^\mathrm{0}_{\mathrm{series}}$, $n_\mathrm{0}$ and $\mu$ as fitting parameters and the $C_\mathrm{g}$ calculated as $\frac{(\epsilon\times\epsilon_\mathrm{0})}{d_\mathrm{g}}$, where the $\epsilon_\mathrm{0}$ is the vacuum permittivity and $\epsilon$ is the dielectric constant of the gate dielectric. We assume that in our GFETs $\epsilon\approx7.5$.\cite{Groner2002} The values of the fitting parameters and the $C_\mathrm{g}$ are given in Table~\ref{tab:fitting1}, where $C_g$ is equivalent to the oxide capacitance ($C_{\mathrm{ox}}$), since no other capacitance is considered. In both GFET-1 and GFET-2 the mobility found using the drain resistance model is 2-3 times lower than the corresponding gMR mobility. We explain it by the limitations of the drain resistance model, which includes the assumption of constant mobility and uncertainty of the gate capacitance. In particular, in our previous studies, using the delay time analysis and capacitance-voltage characteristics, we showed explicitly that the drain resistance model, in its commonly used approach, i.e., using only $C_{\mathrm{ox}}$ as the gate capacitance, can underestimate the mobility. Furthermore, we have shown that the gate capacitance in GFETs can be strongly modified by the interfacial states, which further introduces uncertainties in the capacitance value.~\cite{Bonmann2017,bonmann2017olof} In summary, the gMR method of extraction of mobility can be considered as potentially more accurate, in comparison with the commonly used approach of fitting by the drain resistance model, since it does not require the assumption of the constant mobility and knowledge of the gate capacitance, i.e., it is free from these limitations and associated uncertainties.
        \begin{table}[]
            \centering
            \caption{Parameters used in the drain resistance model.}
            \begin{tabular}{l|c|c|c|c}
                       & $\mu$~(cm$^2$/Vs) & $R_{0}$~($\Omega$) & $n_0\cdot10^{15}$~(m$^{-2}$) & $C_{ox}\cdot10^{-3}$~(F/m$^2$)\\\hline
                GFET-1  &\cellcolor{blue!15}  868 &\cellcolor{blue!15} 2150 &\cellcolor{blue!15} 10 & 3 \\\hline
                GFET-2  &\cellcolor{blue!15} 1380 &\cellcolor{blue!15}   40 &\cellcolor{blue!15} 10 & 3  \\\hline
                        &  \multicolumn{3}{c|}{\cellcolor{blue!15} fitting parameters} & 
            \end{tabular}
            \label{tab:fitting1}
        \end{table}
\par
    In conclusion, we show that the geometrical magnetoresistance effect can be used to extract and study mobility directly in the graphene field-effect transistors, i.e., without involving additional specific test structures. This allows for avoiding the significant uncertainties associated with the strong surface distribution of the mobility over the wafer surface caused by the spatially inhomogeneous Coulomb potential of charged impurities. In contrast to the commonly used approach of fitting by the drain resistance model, the gMR method does not require the assumption of the constant mobility and knowledge of gate capacitance, and is, therefore, free from the limitations and potentially more accurate. Furthermore, the gMR method allows for studying and interpreting of the measured dependencies of mobility on the gate voltage, i.e., carrier concentration, and identifying the corresponding scattering mechanisms. In particular, the decrease in mobility at higher carrier concentrations observed in this work is associated with the contribution of the phonon scattering. Finally, we show that the gMR method is applicable even to the GFETs with relatively high series resistance with a specific width-resistivity up to 5.5~k$\Omega\times\mu$m.

\subsection*{Supplementary Material}
    Details of a thorough error analysis of the data, and a description of the measurements set-up, can be found in the supplementary material.

\subsubsection*{Acknowledgment}
    This work was supported in part by the EU Graphene Flagship Core 3 Project under Grant 881603. This work was performed in part at Myfab Chalmers and in part at Micronova Nanofabrication Centre. The authors thank Saroj Prasad Dash for assisting us with the measurement laboratory. AG and SA acknowledge funding from the Academy of Finland (grant nos. 343842 and 314809), AM and MS acknowledge funding from the Graphene Flagship 2D Experimental Pilot Line. The authors also thank Profs. Jan Stake and Kjell Jeppson for fruitful discussions and Prof. Niklas Rorsman for help with measurements, constructive feedback during discussions and the writing process. The authors also thank Muhammas Asad for his assistance in the GFET design and fabrication techniques.

\bibliography{main}

%

\section*{Supplementary Information}\label{SupInf}\label{sec:sup}
    \subsection{Characterization of GFETs}\label{sup:a}
\begin{figure}[h!]
                \centering
                \includegraphics[width=\columnwidth]{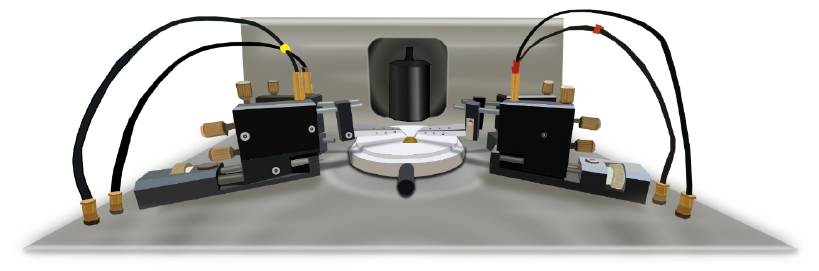}
                \caption{Illustration of the experimental set-up with a permanent magnet visible in the middle, used for measurements of the GFET transfer characteristics.}
                \label{fig:setup}
            \end{figure}

        The transfer characteristics of the GFETs were measured using dc microprobes in a standard Hall set-up \emph{BIO-RAD HL5500PC} equipped by a movable permanent magnet with $B=0.33$~T, as depicted in Fig.~\ref{fig:setup}.
            
    \subsection{Error analysis}\label{SupInf:DUT1}\label{sup:b}
        We assume, that the error in the gMR mobility is defined mainly by errors in the resistances of the gated regions of the GFET channel and that the errors with and without magnetic field are equal, i.e. $\delta R^B_{\mathrm{gated}} = \delta R^0_{\mathrm{gated}}$. Uncertainty in the magnetic field is assumed to be relatively low. The $\delta R^0_{\mathrm{gated}}$ is composed by the errors in the measured drain resistance ($\delta R^0_{\mathrm{total}}$) and the series resistance ($\delta R^0_{\mathrm{series}}$) and assuming propagation of errors can be expressed as 
            \begin{equation}\label{eq:propErrorResistance}\tag{S1}
                \delta R^0_{\mathrm{gated}} = \sqrt{(\delta R^0_{\mathrm{total}})^2 + (\delta R^0_{\mathrm{series}})^2}.
            \end{equation}
\par
        The error in the drain resistance is caused by the drain current instabilities due to trapping/de-trapping of the charge carriers. The error in the series resistance is caused by uncertainty in the linear fitting by a software algorithm. It can be shown, that, using Eq: $\mu_{\mathrm{gMR}}\approx\frac{1}{B}\sqrt{\frac{R^B_{\mathrm{gated}}}{R^0_{\mathrm{gated}}}-1}$ for geometrical magnetoresistance mobility and propagation of errors, the relative error in the gMR mobility can be expressed as 
            \begin{gather}\label{eq:propErrorMobility}\tag{S2}
                \frac{\delta\mu_{\mathrm{gMR}}}{\mu_{\mathrm{gMR}}} = \frac{\delta R^0_{\mathrm{gated}}}{R^0_{\mathrm{gated}}}f\big(\frac{R^B_{\mathrm{gated}}}{R^0_{\mathrm{gated}}}\big) \\  \approx\frac{\delta R^0_{\mathrm{gated}}}{2(R^B_{gated} - R^0_{\mathrm{gated}})}\sqrt{1+\Big(\frac{R^B_{\mathrm{gated}}}{R^0_{\mathrm{gated}}}\Big)^2}.\tag{S3}\label{eq:propErrorMobilityS3}
            \end{gather}
\par
        Given below are errors calculated for GFET-1 and GFET-2. We use standard errors for the $\delta R^0_{\mathrm{total}}$ and $\delta R_{\mathrm{series}}^0$. The data and the linear fitting shown for GFET-2 give $\delta R^0_{\mathrm{series}} = 0.22\ \Omega$. The $\delta R_{\mathrm{total}}$ values for GFET-1 are calculated from standard deviations via statistical analysis of 60 gate voltage sweeps and found to be below 0.2~$\%$ in the whole range of the gate voltage. Calculations using Eqs.~(\ref{eq:propErrorResistance}), (\ref{eq:propErrorMobility}) and (\ref{eq:propErrorMobilityS3}) show that the relative error in the gMR mobility is below 8~$\%$ in the whole range of the gate voltage.
        



\end{document}